\documentclass[letterpaper,10pt,xcolor=tex,dvipsnames]{article} 

\newcommand{\mb}{\mathbf}
\usepackage{multirow}
\usepackage{wrapfig}
\DeclareMathAlphabet{\pazocal}{OMS}{zplm}{m}{n}
\SetMathAlphabet\pazocal{bold}{OMS}{zplm}{bx}{n}

\usepackage{osameet3} 

\usepackage{amsmath,amssymb}
\usepackage[colorlinks=true,bookmarks=false,citecolor=blue,urlcolor=blue]{hyperref} 
\usepackage{subcaption}
\usepackage{pgfplots,tikz}
\usetikzlibrary{shapes}
\pgfplotsset{compat = newest}

\definecolor{mycolor1}{rgb}{0.85490,0.92941,0.70196}%
\definecolor{mycolor2}{rgb}{0.45098,0.27451,0.94902}%
\definecolor{mycolor3}{rgb}{0.61569,0.24706,0.20000}%
\definecolor{mycolor4}{rgb}{0.94510,0.14510,0.73725}%
\definecolor{mycolor5}{rgb}{0.66275,0.81569,0.96078}%
\definecolor{mycolor6}{rgb}{0.14118,0.90980,0.24706}%

\begin{document}
\title{Model-aided Geometrical Shaping of Dual-polarization 4D Formats in the Nonlinear Fiber Channel}

\author{Gabriele Liga$^1$, Bin Chen$^{1,2}$, and Alex Alvarado$^1$}
\address{$^1$Department of Electrical Engineering, Eindhoven University of Technology, Eindhoven, The Netherlands\\
$^2$School of Computer Science and Information Engineering, Hefei University of Technology, Hefei, China}
\email{g.liga@tue.nl}

\copyrightyear{2021}

\begin{abstract}
The geometry of dual-polarization four-dimensional constellations is optimized in the optical fiber channel using a recent nonlinear interference model. 
A 0.27 bit/4D rate gain and 13\% reach increase are attained compared to polarization-multiplexed formats. 
\end{abstract}

\section{Introduction}
\vspace{-.2cm}
Multidimensional constellation shaping is an effective approach to harvest spectral efficiency gains in the optical fiber channel. In the additive white Gaussian noise channel (AWGN), higher gains are to be expected from geometrical shaping when increasing the number of dimensions of the constellation. However, unlike the AWGN channel where all channel degrees of freedom are alike, in the nonlinear fiber channel shaping gains depend not only on the dimensionality of the transmitted constellation but also on the specific channel degrees of freedoms used for transmission. As an example, dual-polarization four-dimensional (DP-4D) formats, i.e.~4D constellations that are mapped onto the four available degrees of freedom of the dual-polarization optical field, have been recently rediscovered for their increased nonlinearity tolerance as well as increased achievable information rates compared to polarization-multiplexed 2D (PM-2D) formats \cite{Kojima2017, Chen2019, Chen2021}.

Constellation optimization in the nonlinear fiber channel is typically performed numerically using the split-step Fourier channel or via nonlinear interference (NLI) power models. In particular, NLI power models allow the quick computation of an effective signal-to-noise ratio (SNR) as a function of the input constellation, which is required for the computation of the objective function to optimize. However, almost all available NLI models only apply to PM-2D modulation formats is used. This limitation prevents an accurate model-based optimization of the transmitted constellation in the full DP-4D space. To circumvent this obstacle, previous works have made use of heuristic ideas, such as the constant-modulus constraint, to design good nonlinearity-tolerant constellations in 4D \cite{Chen2019,Chen2021}, whilst maximising the linear shaping gain. 

Recently in \cite{Liga2020}, we have introduced a new analytical expression for the prediction of the NLI power when a general DP-4D linear modulation is adopted. The model expresses the dependency of the NLI power on the modulation format in closed-form thus enabling a fast search of the most nonlinearly tolerant DP-4D constellations. In particular, in \cite{liga2021OFC}, we showed how in the DP-4D space, constant-modulus constellations are not necessarily the most robust against nonlinear distortion, and as as result, the problem of finding the optimal DP-4D constellations for optical fiber transmission is still open. 

In this work, we present a first numerical optimization of the achievable information rate (AIR) for geometrically-shaped DP-4D formats supported by the use of an NLI analytical model. A set of four novel 4D constellations which outperform previously known 4D formats is presented. The results in this work confirm once again the potential of DP-4D constellations in fiber-optic transmission and leave room for further improvements.  
\section{Methodology}
\vspace{-.2cm}
In a channel with $N$-dimensional input $\mb{x}$, output $\mb{y}$, and marginal distribution $p(\mb{y}|\mb{x})$, the quantity
\begin{equation}
I=\log_2M+\frac{1}{M}\sum_{\mb{x}\in \pazocal{X}}\int_{\mathbb{R}^{N}} p(\mb{y}|\mb{x})\log_2\frac{q(\mb{y}|\mb{x})}{\sum_{\mb{x}^{\prime}\in \pazocal{X}}q(\mb{y}|\mb{x}^{\prime})} d\mb{y}  
\label{eq:mis_AIR}    
\end{equation}
where $M=|\pazocal{X}|$,
is an AIR for the discrete uniformly-distributed multidimensional constellation $\pazocal{X}$ under decoding metric $q(\mb{y}|\mb{x})$ \cite{Arnold2006}. A typical assumption in optical communications is $p(\mb{y}|\mb{x})=q(\mb{y}|\mb{x})$ and both are assumed to be a multivariate Gaussian distribution with diagonal covariance matrix and identical diagonal terms. This assumption allows an analytical computation of \eqref{eq:mis_AIR}, as long as the dependence of the NLI power on the input constellation and power is known. For PM-2D constellations such relationship is provided, e.g., by the enhanced Gaussian noise (EGN) model \cite{Carena2014}. Although the EGN model does not explicitly assume $p(\mb{y}|\mb{x})$ to be Gaussian its NLI power expression can still be used to compute \eqref{eq:mis_AIR} under a circularly-symmetric Gaussian assumption, as done for instance in \cite{SillekensECOC2018}.
In this work, we extend this approach by: i) replacing the expression for the EGN NLI power with the one obtained in \cite[Th.~2]{Liga2020} for DP-4D formats; ii) using a \emph{non-circularly symmetric} 4D Gaussian distribution with diagonal covariance matrix for $q(\mb{y}|\mb{x})$. The latter step is justified by the fact that our 4D NLI model predicts in general non identical NLI powers over the two polarization channels when non PM-2D formats are transmitted.

The numerical optimization of \eqref{eq:mis_AIR} for $N=4$, was performed using a sequential quadratic programming algorithm which is a gradient-based algorithm for optimization problems with a nonlinear constraint. In our case, this constraint is the result of imposing that the (4D) SNR is equal to the optimum SNR achievable by the trial constellation, and it fully defines the covariance matrix of $q(\mb{y}|\mb{x})$. 
\begin{wraptable}{r}{0.335\textwidth}\vspace{-.44cm}
\begin{footnotesize}
\begin{tabular}{c|c}
\hline\hline
\multicolumn{2}{c}{\textbf{TX Parameters}} \\
\hline
Symbol rate & 50 Gbaud \\
No. of channels & 1 \\
Root-raised-cosine roll-off & 0.01\% \\
\hline
\multicolumn{2}{c}{\textbf{Fiber Parameters}} \\
\hline
Attenuation coeff. ($\alpha$) & 0.2 dB/km \\
Disp. parameter ($D$) & 17 ps/nm/km \\
Nonlinear coeff. ($\gamma$) & 1.2 dB/km \\
\hline
\multicolumn{2}{c}{\textbf{Link Parameters}} \\
\hline
Span length & 100 km \\
EDFA noise figure & 5 dB \\
\hline\hline
\end{tabular}
\end{footnotesize}
\vspace{-.38cm}\caption{Numerical study parameters.}\label{tab:opt_const}
\end{wraptable} 
This approach, enabled by the fast model-based computation of the DP-4D constellation NLI vector, avoids a joint optimization of the transmitted power and is justified by the monotonic behaviour of \eqref{eq:mis_AIR} as a function of the effective SNR (for a fixed constellation). To compute \eqref{eq:mis_AIR}, the Gauss-Hermite quadrature with 8 points per dimension was adopted. Results were then validated against split-step Fourier method (SSFM) simulations, where the integral in \eqref{eq:mis_AIR} is computed via a Monte-Carlo (MC) approach using samples of $p(\mb{y}|\mb{x})$ from the SSFM channel. The 4D constellation geometry was optimized for a single-channel, multi-span transmission system with standard single-mode fiber and Erbium-doped fiber amplifiers (EDFAs). The parameters of this system are listed in Table \ref{tab:opt_const}. 
\begin{figure}[!t]
\centering
\begin{subfigure}[b]{0.48\textwidth}
\begin{tikzpicture}

\begin{axis}[%
width=2.7in,
height=1.5in,
scale only axis,
xmin=3000,
xmax=12500,
ymin=4.3,
ymax=6,
ylabel={$I$ [bit/4D]},
ylabel shift=-4pt,
xtick={4000,6000,8000,10000,12000},
ytick={4,4.5,5,5.5,6},
scaled x ticks=false,
axis background/.style={fill=white},
xmajorgrids,
ymajorgrids,
legend style={legend cell align=left, align=left, draw=white!15!black, font=\footnotesize},
every tick label/.append style={font=\small}
]

\addplot [color=red, line width=.6, mark=*,mark size=1.8pt]
  table[row sep=crcr]{%
1000	5.99999835147695\\
2000	5.99891555264508\\
3000	5.98411031412945\\
3500	5.96402216811768\\
4000	5.93439772091315\\
4500	5.88809412813187\\
5000	5.83250409374757\\
5500	5.76424800568287\\
6000	5.69067404209035\\
6500	5.60639930171353\\
7000	5.51532673775764\\
7500	5.42315820405758\\
8000	5.32079542614793\\
8500	5.22056202410218\\
9000	5.1282859120449\\
9500	5.03215861401886\\
10000	4.93327647123731\\
10500	4.83798039072482\\
11000	4.74632543619324\\
11500	4.63925689009331\\
12000	4.54717400751225\\
13000	4.37623506938401\\
14000	4.19461666573109\\
15000	4.04919682464151\\
};
\addlegendentry{$\pazocal{C}^{*}_{64}$}

\addplot [color=red, line width=.6,mark=triangle*,mark options={fill=white,solid},mark size=2.5pt]
  table[row sep=crcr]{%
3000	5.98602531212498\\
3500	5.96567382334538\\
4000	5.92996517697172\\
4500	5.88465115295351\\
5000	5.81987262068359\\
5500	5.75070341602608\\
6000	5.66992664055136\\
6500	5.5780686263975\\
7000	5.48118788656882\\
7500	5.38643975287785\\
8000	5.28574821091847\\
8500	5.18180832419265\\
9000	5.09477243813076\\
9500	4.97976185946594\\
10000	4.88015067350299\\
10500	4.78059733555165\\
11000	4.6902582920915\\
11500	4.589119259361\\
12000	4.50074011601972\\
13000	4.31382915639973\\
14000	4.15986666193342\\
15000	3.99202234045533\\
};
\addlegendentry{w4$\textunderscore$64}

\addplot [color=red, line width=.6, mark=diamond*,mark options={fill=white},mark size=2.5pt]
  table[row sep=crcr]{%
3000	5.9631178660264\\
3500	5.92833890496131\\
4000	5.87948013312732\\
4500	5.81849349433065\\
5000	5.74029623249451\\
5500	5.66577537735492\\
6000	5.57383191086128\\
6500	5.48021793727833\\
7000	5.38853027686078\\
7500	5.2908242451762\\
8000	5.18852857551173\\
8500	5.08944411930073\\
9000	4.99295336533659\\
9500	4.8951156533675\\
10000	4.80544254824695\\
10500	4.70901999337936\\
11000	4.62124285944004\\
11500	4.52642516997818\\
12000	4.44363082456779\\
13000	4.26797430342343\\
14000	4.09744740556253\\
15000	3.95856551403168\\
};
\addlegendentry{4D-64PRS}

\addplot [color=red, line width=.6, mark=square*,mark options={fill=white, solid},mark size=1.7pt]
  table[row sep=crcr]{%
3000	5.95553901807613\\
3500	5.91201126178548\\
4000	5.84980133562511\\
4500	5.77986274279835\\
5000	5.69440168207815\\
5500	5.60539362377674\\
6000	5.50123740994357\\
6500	5.4035814022618\\
7000	5.29667063559475\\
7500	5.19597687863713\\
8000	5.09226432981635\\
8500	4.99516924738682\\
9000	4.89204464917586\\
9500	4.7954165577357\\
10000	4.69156480366114\\
10500	4.60152676139696\\
11000	4.49474712205088\\
11500	4.41605980619586\\
12000	4.33878915742314\\
13000	4.17862813505389\\
14000	4.01046748051678\\
15000	3.86276014931065\\
};
\addlegendentry{PM-8QAM}

\path[fill=yellow,opacity=0.5]  (9500,	4.7954165577357) -- (11500,	4.4160) -- (11500,	4.6392) --(9500,5.0321) -- cycle;
\draw[->,thick](8000,4.9)--(10000,4.6);

\node[font=\footnotesize] at (axis cs:10800, 5.1) (fec){80\% FEC rate};
\draw[->,thick](fec)--(10800,4.8);

\node[] at (axis cs:3400, 5.65) {\textbf{(a)}};
\end{axis}

\begin{axis}[%
width=1.2 in,
height=.75 in,
scale only axis,
at={(4.6,10)},
xtick={10000,11000},
scaled x ticks=false,
xmin=9400,
xmax=11700,
ymin=4.5,
ymax=5,
ytick={4.6,4.8},
xticklabel style = {yshift=0.04cm},
axis background/.style={fill=white},
xmajorgrids,
ymajorgrids,
every tick label/.append style={font=\small}
]

\draw[<->,thick] (axis cs:9500,4.8)--++(31pt,0pt)node[above]{$+13\%$}--(axis cs:10700,4.8);
\draw[<->,thick] (axis cs:10700,4.56)--++(0pt,13pt)node[right,xshift=-1pt]{$0.24$ bit}--(axis cs:10700,4.8);

\addplot [color=red, line width=.6, mark=*,mark size=1.8pt]
  table[row sep=crcr]{%
1000	5.99999835147695\\
2000	5.99891555264508\\
3000	5.98411031412945\\
4000	5.93439772091315\\
5000	5.83250409374757\\
6000	5.69067404209035\\
7000	5.51532673775764\\
8000	5.32079542614793\\
9000	5.1282859120449\\
10000	4.93327647123731\\
11000	4.74632543619324\\
12000	4.54717400751225\\
13000	4.37623506938401\\
14000	4.19461666573109\\
15000	4.04919682464151\\
};

\addplot [color=red, line width=.6,mark=triangle*,mark options={fill=white,solid},mark size=2.5pt]
  table[row sep=crcr]{%
3000	5.98602531212498\\
4000	5.92996517697172\\
5000	5.81987262068359\\
6000	5.66992664055136\\
7000	5.48118788656882\\
8000	5.28574821091847\\
9000	5.09477243813076\\
10000	4.88015067350299\\
11000	4.6902582920915\\
12000	4.50074011601972\\
13000	4.31382915639973\\
14000	4.15986666193342\\
15000	3.99202234045533\\
};

\addplot [color=red, line width=.6, mark=diamond*,mark options={fill=white},mark size=2.5pt]
  table[row sep=crcr]{%
3000	5.9631178660264\\
4000	5.87948013312732\\
5000	5.74029623249451\\
6000	5.57383191086128\\
7000	5.38853027686078\\
8000	5.18852857551173\\
9000	4.99295336533659\\
10000	4.80544254824695\\
11000	4.62124285944004\\
12000	4.44363082456779\\
13000	4.26797430342343\\
14000	4.09744740556253\\
15000	3.95856551403168\\
};

\addplot [color=red, line width=.6, mark=square*,mark options={fill=white, solid},mark size=1.7pt]
  table[row sep=crcr]{%
3000	5.95553901807613\\
4000	5.84980133562511\\
5000	5.69440168207815\\
6000	5.50123740994357\\
7000	5.29667063559475\\
8000	5.09226432981635\\
9000	4.89204464917586\\
10000	4.69156480366114\\
11000	4.49474712205088\\
12000	4.33878915742314\\
13000	4.17862813505389\\
14000	4.01046748051678\\
15000	3.86276014931065\\
};

\end{axis}

\end{tikzpicture}%
\label{subfig:a}
\end{subfigure}\hfill
\begin{subfigure}[b]{0.475\textwidth}
\begin{tikzpicture}
\begin{axis}[%
width=2.7in,
height=1.5in,
scale only axis,
xmin=1800,
xmax=9000,
ymin=5,
ymax=7,
xtick={2000,4000,6000,8000,10000},
ylabel shift=-4pt,
ytick={5,5.5,6,6.5,7},
scaled x ticks=false,
axis background/.style={fill=white},
xmajorgrids,
ymajorgrids,
legend style={legend cell align=left, align=left, draw=white!15!black, font=\footnotesize},
every tick label/.append style={font=\small}
]
\addplot [color=blue, line width=.8, mark=*]
  table[row sep=crcr]{%
2000	6.98675632221277\\
2500	6.9563455393815\\
3000	6.90255399684607\\
3500	6.82079452470643\\
4000	6.72156994175682\\
4500	6.60084778093286\\
5000	6.47904187699487\\
5500	6.33809529324365\\
6000	6.19564765016821\\
6500	6.0574310870007\\
7000	5.90981999426892\\
7500	5.75383446869199\\
8000	5.62309379461545\\
9000	5.35071893484631\\
10000	5.10919934012386\\
11000	4.86946559774992\\
12000	4.64260568479401\\
};
\addlegendentry{$\pazocal{C}^{*}_{128}$}

\addplot [color=blue, line width=.8, mark=triangle*,mark options={fill=white}]
  table[row sep=crcr]{%
2000	6.98981680752815\\
2500	6.96065748534479\\
3000	6.90407163284436\\
3500	6.81468516522702\\
4000	6.7100102977295\\
4500	6.58308899795861\\
5000	6.43496426904688\\
5500	6.29140971801672\\
6000	6.13140489428291\\
6500	5.9783243132643\\
7000	5.83599165589299\\
7500	5.69153308106308\\
8000	5.54233869212211\\
9000	5.27228740014287\\
10000	5.02777884983933\\
11000	4.77315699329565\\
12000	4.57307535085873\\
};
\addlegendentry{l4$\textunderscore$128}

\addplot [color=blue, line width=.8, mark=square*,mark options={fill=white}]
  table[row sep=crcr]{%
2000	6.90146679550833\\
2500	6.8324402180718\\
3000	6.74927799235057\\
3500	6.64706920448107\\
4000	6.52790119975282\\
4500	6.40739690578237\\
5000	6.27708819614246\\
5500	6.15126805418592\\
6000	6.01718894113311\\
6500	5.88091674495821\\
7000	5.74877628768481\\
7500	5.61508668935626\\
8000	5.48743874149283\\
9000	5.23147783958712\\
10000	5.01143174505395\\
11000	4.7890411401443\\
};

\addlegendentry{4D-OS128}

\path[fill=yellow,opacity=0.5]  (7500,	5.615) -- (7500, 5.7538) --(8500,5.480)--(8500,5.361) -- cycle;
\draw[->,thick](5500,5.8)--(7700,5.5);

\node[font=\footnotesize] at (axis cs:8000, 6.05) (fec){80\% FEC rate};
\draw[->,thick](fec)--(8000,5.7);

\node[] at (axis cs:2200, 6.6) {\textbf{(b)}};
\end{axis}

\begin{axis}[%
width=1.2 in,
height=.8 in,
scale only axis,
at={(2.3,3.5)},
xtick={8000,9000, 10000},
scaled x ticks=false,
xmin=7400,
xmax=8600,
ymin=5.45,
ymax=5.65,
ytick={5.5,5.6},
axis background/.style={fill=white},
xmajorgrids,
ymajorgrids,
every tick label/.append style={font=\small}
]

\addplot [color=blue, line width=.8, mark=*, mark repeat=10]
  table[row sep=crcr]{%
2000	6.98675632221277\\
2500	6.9563455393815\\
3000	6.90255399684607\\
3500	6.82079452470643\\
4000	6.72156994175682\\
4500	6.60084778093286\\
5000	6.47904187699487\\
5500	6.33809529324365\\
6000	6.19564765016821\\
6500	6.0574310870007\\
7000	5.90981999426892\\
7500	5.75383446869199\\
8000	5.62309379461545\\
9000	5.35071893484631\\
10000	5.10919934012386\\
11000	4.86946559774992\\
12000	4.64260568479401\\
};

\addplot [color=blue, line width=.8, mark=triangle*,mark options={fill=white}, mark repeat=10]
  table[row sep=crcr]{%
2000	6.98981680752815\\
2500	6.96065748534479\\
3000	6.90596177203979\\
3000	6.90407163284436\\
3500	6.81468516522702\\
4000	6.71032968725956\\
4000	6.7100102977295\\
4500	6.58308899795861\\
5000	6.43428553356995\\
5000	6.43496426904688\\
5500	6.29140971801672\\
6000	6.13140489428291\\
6500	5.9783243132643\\
7000	5.82733230576842\\
7500	5.69153308106308\\
8000	5.54233869212211\\
9000	5.27228740014287\\
10000	5.02777884983933\\
11000	4.77315699329565\\
12000	4.57307535085873\\
};

\addplot [color=blue, line width=.8, mark=square*,mark options={fill=white}, mark repeat=10]
  table[row sep=crcr]{%
2000	6.90146679550833\\
2500	6.8324402180718\\
3000	6.74927799235057\\
3500	6.64706920448107\\
4000	6.52790119975282\\
4500	6.40739690578237\\
5000	6.27708819614246\\
5500	6.15126805418592\\
6000	6.01718894113311\\
6500	5.88091674495821\\
7000	5.74877628768481\\
7500	5.61508668935626\\
8000	5.48743874149283\\
9000	5.23147783958712\\
10000	5.01143174505395\\
11000	4.7890411401443\\
};

\draw[<->,thick] (axis cs:7550,5.6)--++(20pt,0pt)node[above]{$+7\%$}--(axis cs:8100,5.6);
\draw[<->,thick] (axis cs:8100,5.46)--++(0pt,20pt)node[right,xshift=-3pt]{$0.14$ bit}--(axis cs:8100,5.6);

\end{axis}

\end{tikzpicture}%
\label{subfig:b}
\end{subfigure}\\
\vspace{-.55cm}
\begin{subfigure}[b]{0.47\textwidth}
\begin{tikzpicture}

\begin{axis}[%
width=2.7in,
height=1.5in,
scale only axis,
xmin=1000,
xmax=8000,
ytick={5.5,6,6.5,7,7.5,8},
ymin=5.2,
ymax=8,
ylabel={$I$ [bit/4D]},
xlabel={Distance [km]},
xlabel shift=-5pt,
xtick={1000,2000,4000,6000,8000},
scaled x ticks=false,
axis background/.style={fill=white},
xmajorgrids,
ymajorgrids,
legend style={legend cell align=left, align=left, draw=white!15!black,font=\footnotesize},
every tick label/.append style={font=\small}
]
\addplot [color=magenta,line width=.8, mark=*]
  table[row sep=crcr]{%
1000	7.99765340516049\\
1500	7.97573923234374\\
2000	7.9111008138196\\
2500	7.79890300562121\\
3000	7.64397246207934\\
3500	7.45827370485049\\
4000	7.25941860861286\\
4500	7.04934190008197\\
5000	6.8568488659503\\
5500	6.64077909785929\\
6000	6.4547829290679\\
6500	6.25922062827107\\
7000	6.0816977296367\\
7500	5.9095083338728\\
8000	5.74237112823567\\
9000	5.44450724258898\\
10000	5.15980253803281\\
11000	4.90071574444784\\
12000	4.68660023459416\\
13000	4.46877856845779\\
14000	4.26333505345111\\
15000	4.09408951887377\\
};

\addlegendentry{$\pazocal{C}^{*}_{256}$}

\addplot [color=magenta, line width=.8, mark=triangle*,mark options={fill=white, mark size=3}]
  table[row sep=crcr]{%
1000	7.99946360761338\\
1500	7.98632365672803\\
2000	7.92978980554539\\
2500	7.81474981169903\\
3000	7.65212684852578\\
3500	7.45318880707877\\
4000	7.2275671159059\\
4500	7.00277073765899\\
5000	6.78871953486257\\
5500	6.57349487003972\\
6000	6.36868640939878\\
6500	6.17843514344855\\
7000	6.0030428199107\\
7500	5.82650566454137\\
8000	5.66685609691776\\
};
\addlegendentry{w4$\textunderscore$256}

\addplot [color=magenta, line width=.8, mark=square*,mark options={fill=white, mark size=1.7}]
table[row sep=crcr]{%
1000	7.99527986437082\\
1500	7.95155905200068\\
2000	7.84096952909455\\
2500	7.67408314862646\\
3000	7.47023265492755\\
3500	7.24672625850197\\
4000	7.02782818978518\\
4500	6.80124345350226\\
5000	6.57857428881927\\
5500	6.37827592665697\\
6000	6.17870837517646\\
6500	6.00450024773189\\
7000	5.83586407471139\\
7500	5.67100162119423\\
8000	5.52548029677084\\
};
\addlegendentry{PM-16QAM}

\path[fill=yellow,opacity=0.5]  (5500,	6.37827592665697) -- (5500,	6.64077909785929) -- (6500,	6.25922062827107) --(6500,	6.00450024773189) -- cycle;
\draw[->,thick](4500,6.2)--(5700,6.25);
\node[font=\footnotesize] at (axis cs:6700, 6.8) (fec){80\% FEC rate};
\draw[->,thick](fec)--(6200,6.47);

\node[] at (axis cs:1200, 7.6) {\textbf{(c)}};
\end{axis}

\begin{axis}[%
width=1.2 in,
height=.8 in,
scale only axis,
at={(2.5,8)},
xtick={6000,8000},
scaled x ticks=false,
xmin=5300,
xmax=6700,
ymin=6.1,
ymax=6.5,
ytick={6.2,6.4},
axis background/.style={fill=white},
xmajorgrids,
ymajorgrids,
every tick label/.append style={font=\small}
]

\addplot [color=magenta,line width=.8, mark=*,mark options={fill=white}]
  table[row sep=crcr]{%
1000	7.99765340516049\\
2000	7.9111008138196\\
3000	7.64397246207934\\
4000	7.25941860861286\\
5000	6.8568488659503\\
6000	6.45557513045066\\
7000	6.0816977296367\\
8000	5.74237112823567\\
9000	5.44450724258898\\
10000	5.15980253803281\\
11000	4.90071574444784\\
12000	4.68660023459416\\
13000	4.46877856845779\\
14000	4.26333505345111\\
15000	4.09408951887377\\
};

\addplot [color=magenta, line width=.8, mark=triangle*,mark options={fill=white, mark size=3}, mark repeat=10]
  table[row sep=crcr]{%
1000	7.99946360761338\\
1500	7.98632365672803\\
2000	7.92978980554539\\
2500	7.81474981169903\\
3000	7.65212684852578\\
3500	7.45318880707877\\
4000	7.2275671159059\\
4500	7.00277073765899\\
5000	6.78871953486257\\
5500	6.57349487003972\\
6000	6.36868640939878\\
6500	6.17843514344855\\
7000	6.0030428199107\\
7500	5.82650566454137\\
8000	5.66685609691776\\
};

\addplot [color=magenta, line width=.8, mark=square*,mark options={fill=white, mark size=1.7}, mark repeat=10]
table[row sep=crcr]{%
1000	7.99527986437082\\
1500	7.95155905200068\\
2000	7.84096952909455\\
2500	7.67408314862646\\
3000	7.47023265492755\\
3500	7.24672625850197\\
4000	7.02782818978518\\
4500	6.80124345350226\\
5000	6.57857428881927\\
5500	6.37827592665697\\
6000	6.17870837517646\\
6500	6.00450024773189\\
7000	5.83586407471139\\
7500	5.67100162119423\\
8000	5.52548029677084\\
};

\draw[<->,thick] (axis cs:5450,6.4)--++(10pt,0pt)node[above]{$+13\%$}--(axis cs:6150,6.4);
\draw[<->,thick] (axis cs:6150,6.13)--++(0pt,21pt)node[right,xshift=-2pt]{$0.27$ bit}--(axis cs:6150,6.4);
\end{axis}

\end{tikzpicture}%
\label{subfig:c}
\end{subfigure}
\hfill\begin{subfigure}[b]{0.47\textwidth}
%
%
\definecolor{mycolor1}{rgb}{0.01569,0.44706,0.37255}%
\definecolor{mycolor2}{rgb}{0.75686,0.10588,0.21176}%
\definecolor{mycolor3}{rgb}{0.91765,0.86667,0.45490}%
\definecolor{mycolor4}{rgb}{0.92941,0.08627,0.65098}%
\definecolor{mycolor5}{rgb}{0.18824,0.53333,0.50588}%
\begin{tikzpicture}

\begin{axis}[%
width=2.7in,
height=1.5in,
scale only axis,
xlabel={Distance [km]},
xlabel shift=-5pt,
xtick={2000,3000,4000,5000},
scaled x ticks=false,
xmin=1000,
xmax=5500,
ymin=6,
ymax=9,
ytick={6,6.5,7,7.5,8,8.5,9},
axis background/.style={fill=white},
xmajorgrids,
ymajorgrids,
legend style={legend cell align=left, align=left, draw=white!15!black,font=\footnotesize},
every tick label/.append style={font=\small}
]
\addplot [color=Green,line width=.8, mark=*]
  table[row sep=crcr]{%
1000	8.98137194260343\\
1500	8.88462832230112\\
2000	8.69430317288206\\
2500	8.44444295109541\\
3000	8.15678197384955\\
3500	7.85548981575069\\
4000	7.57707018329349\\
4500	7.29173811041976\\
5000	7.0285914211717\\
5500	6.77765953246191\\
6000	6.56359790083486\\
6500	6.3367065909837\\
7000	6.15268961115936\\
7500	5.969347451759\\
8000	5.78371201067012\\
8500	5.62835846394866\\
9000	5.47188512218626\\
9500	5.31970591129526\\
10000	5.18553565767905\\
};
\addlegendentry{$\pazocal{C}^{*}_{512}$}

\addplot [color=Green,line width=.8, mark=square*,mark options={fill=white}]
  table[row sep=crcr]{%
1000	8.9912227883386\\
1500	8.91013610695655\\
2000	8.7139124729794\\
2500	8.43074300354348\\
3000	8.11811488567057\\
3500	7.79218717347819\\
4000	7.48996050794179\\
4500	7.20080030817327\\
5000	6.93962068584028\\
6000	6.46533510816072\\
7000	6.06291624096438\\
8000	5.7095777085187\\
};
\addlegendentry{a4$\textunderscore$512}
\node[] at (axis cs:1200, 8.59) {\textbf{(d)}};

\path[fill=yellow]  (4500, 7.20080030817327) -- (4500,	7.29173811041976) -- (5000,	7.0285914211717) -- (5000,	6.93962068584028) -- cycle;
\draw[->,thick](3000,7.2)--(4700,7.05);
\node[font=\footnotesize] at (axis cs:4800, 7.65) (fec){80\% FEC rate};
\draw[->,thick](fec)--(4800,7.2);
\end{axis}

\begin{axis}[%
width=1.2 in,
height=.8 in,
scale only axis,
at={(7,4)},
xtick={4500,4700,4900},
scaled x ticks=false,
xmin=4500,
xmax=4800,
ymin=7.05,
ymax=7.25,
ytick={7.1,7.2},
axis background/.style={fill=white},
xmajorgrids,
ymajorgrids,
every tick label/.append style={font=\small}
]

\addplot [color=Green,line width=.8, mark=*,mark options={fill=white}]
  table[row sep=crcr]{%
1000	8.98137194260343\\
1500	8.88462832230112\\
2000	8.69430317288206\\
2500	8.44444295109541\\
3000	8.15678197384955\\
3500	7.85548981575069\\
4000	7.57707018329349\\
4500	7.29173811041976\\
5000	7.0285914211717\\
5500	6.77765953246191\\
6000	6.56359790083486\\
6500	6.3367065909837\\
7000	6.15268961115936\\
7500	5.969347451759\\
8000	5.78371201067012\\
8500	5.62835846394866\\
9000	5.47188512218626\\
9500	5.31970591129526\\
10000	5.18553565767905\\
};

\addplot [color=Green,line width=.8, mark=square*,mark options={fill=white}, mark repeat=7]
  table[row sep=crcr]{%
1000	8.9912227883386\\
1500	8.91013610695655\\
2000	8.7139124729794\\
2500	8.43074300354348\\
3000	8.11811488567057\\
3500	7.79218717347819\\
4000	7.48996050794179\\
4500	7.20080030817327\\
5000	6.93962068584028\\
6000	6.46533510816072\\
7000	6.06291624096438\\
8000	5.7095777085187\\
};

\draw[<->,thick] (axis cs:4504,7.2)--++(15pt,0pt)node[above]{$+4$\%}--(axis cs:4670,7.2);
\draw[<->,thick] (axis cs:4670,7.11)--++(0pt,12pt)node[right,xshift=-2pt]{$0.09$ bit}--(axis cs:4670,7.2);
\end{axis}

\end{tikzpicture}%
\label{subfig:d}
\end{subfigure}
\vspace{-.8cm}
\caption{Rate $I$ in \eqref{eq:mis_AIR} vs transmission distance for the optimized DP-4D constellations $\pazocal{C}^*_{M}$ in this paper, with $m=6$ (a), $m=7$ (b), $m=8$ (c), and $m=9$ (d). Other known 4D constellations are also shown as a reference.}
\label{fig:AIRvsDist}
\vspace{-.7cm}
\end{figure}
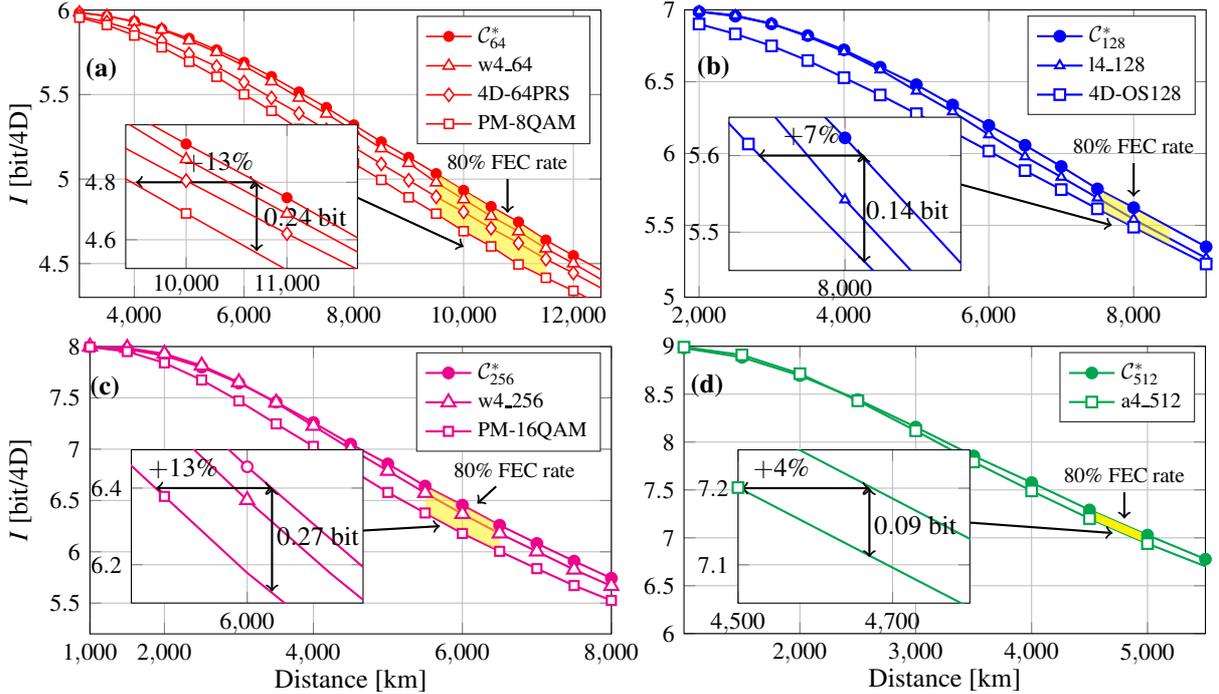

\vspace{.15cm}
\section{Results}
\vspace{-.15cm}
The optimization of the constellation geometry was performed for four different spectral efficiencies $m=\log_2M$ bit/4D with $m\in\{6,7,8,9\}$. The resulting constellations, hereby referred as $\pazocal{C}^*_{M}$ were compared to their corresponding PM-2D formats (for $m\in\{6,8\}$) or with other 4D formats which are known to perform well, either in the AWGN or in the optical fiber channel. Examples of such constellations are, the Welti constellations \cite{Welti1974} or the 4D 64 polarization-ring-switching (4D-64PRS) format \cite{Chen2019}. 

As the result of the optimization was observed to be dependent on the initial constellation, the best 4D format among the ones listed in \cite{AgrellCodesSE} was first identified, and then used as an initial value for the optimization. Namely, the initial constellation was set to the 64 point Welti constellation (w4\textunderscore64), l4\textunderscore128 \cite{AgrellCodesSE}, the 256-point Welti constellation (w4\textunderscore256), and a4\textunderscore512 \cite{AgrellCodesSE} which is a conjectured optimal packing. To target an 80\% forward error correction (FEC) code rate, at each $m$ the optimization was performed at a transmission distance where $I=0.8m$ is achieved by the baseline constellations. The rate $I$ in \eqref{eq:mis_AIR} of the $\pazocal{C}^*_{M}$ constellations was then computed as a function of the transmission distance using a SSFM-based MC approach. The results are shown in Fig.~\ref{fig:AIRvsDist}. 

For $m=6$, $\pazocal{C}^*_{64}$, achieves a 0.24 bit/4D gain and 13\% longer transmission reach compared to PM 8 quadrature amplitude modulation (PM-8QAM) and about 0.12 bit/4D higher rate compared to 4D-64PRS. The performance of w4\textunderscore64 is only marginally lower than $\pazocal{C}^*_{64}$. For $m=7$, $\pazocal{C}^*_{128}$ gains 0.14 bit/4D or 7\% reach increase over the 4D orthant-symmetric constellation in \cite{Chen2021}. This gain reduces to less than 0.1 bit/4D if compared to the l4\textunderscore128 constellation \cite{AgrellCodesSE}. For $m=8$, 0.27 bit/4D higher rate or 13\% reach increase over PM-16QAM is obtained using $\pazocal{C}^*_{256}$. As in the $m=6$ case, the Welti constellation w4\textunderscore256 only loses less than 0.1 bit/4D vs our optimized $\pazocal{C}^*_{256}$. Finally, for $m=9$, our proposed $\pazocal{C}^*_{512}$ achieves only a 0.09 bit/4D higher rate than a4\textunderscore512 with a 4\% reach increase.


\begin{figure}[!t]
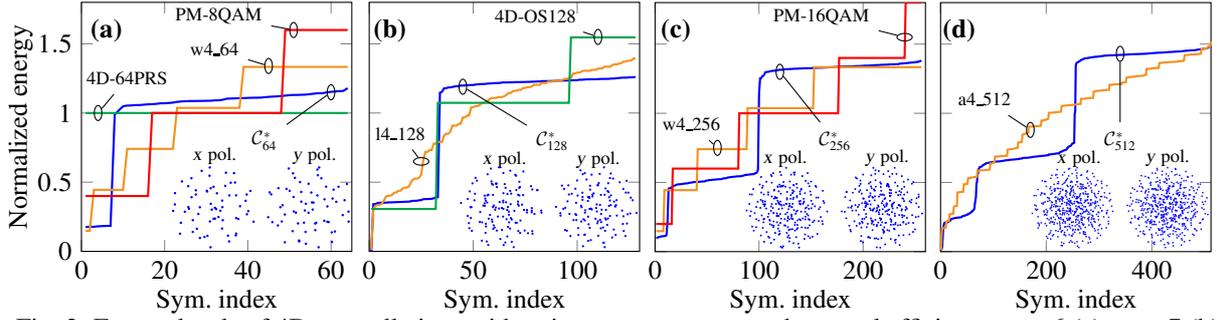

\hspace{-.3cm}
\begin{subfigure}[b]{0.29\textwidth}
%
%
\definecolor{mycolor1}{rgb}{0.00000,0.44700,0.74100}%
\definecolor{mycolor2}{rgb}{0.85000,0.32500,0.09800}%
\definecolor{mycolor3}{rgb}{0.92900,0.69400,0.12500}%
\definecolor{mycolor4}{rgb}{0.49400,0.18400,0.55600}%
\begin{tikzpicture}

\begin{axis}[%
width=1.4in,
height=1.3in,
scale only axis,
xmin=0,
xmax=65,
xlabel={Sym. index},
xlabel shift=-3pt,
ymin=0,
ymax=1.8,
ylabel={Normalized energy},
ylabel shift=-5pt,
axis background/.style={fill=white},
legend pos={south east},
legend style={legend cell align=left, align=left, draw=white!15!black, font=\scriptsize}
]
\addplot [color=blue, line width=.8]
  table[row sep=crcr]{%
1	0.175772685838115\\
2	0.177434747794173\\
3	0.177694796169322\\
4	0.181104125732186\\
5	0.182568550545884\\
6	0.183783430432527\\
7	0.184054860901165\\
8	0.988136371341618\\
9	1.01970451219005\\
10	1.0484166136536\\
11	1.05371867862191\\
12	1.05463776560489\\
13	1.05712128609387\\
14	1.05744973912437\\
15	1.06036949451145\\
16	1.06278343844951\\
17	1.06388396919247\\
18	1.06620806002805\\
19	1.0703017125201\\
20	1.07040638849171\\
21	1.0723858971677\\
22	1.07519053462014\\
23	1.08047038001815\\
24	1.08340908269408\\
25	1.08509493971161\\
26	1.08648492444809\\
27	1.08662317154766\\
28	1.08716527384822\\
29	1.08747433032933\\
30	1.08822687567572\\
31	1.08876791184067\\
32	1.09128986801799\\
33	1.09165179305526\\
34	1.09613248695759\\
35	1.10373114206246\\
36	1.10495154665472\\
37	1.10563069140901\\
38	1.10761495229004\\
39	1.10772784470132\\
40	1.10920190444697\\
41	1.11013852188085\\
42	1.11072144684598\\
43	1.11104285597607\\
44	1.11375721696927\\
45	1.1166961839952\\
46	1.11883998401164\\
47	1.11995587049464\\
48	1.1227313772632\\
49	1.12580756304515\\
50	1.12913630462423\\
51	1.13060024346515\\
52	1.13089570393078\\
53	1.13196913556728\\
54	1.13311841652628\\
55	1.13643151465799\\
56	1.14040612106843\\
57	1.14294848083888\\
58	1.14315845644891\\
59	1.14973256581128\\
60	1.15290518380272\\
61	1.15577339903049\\
62	1.15820356047107\\
63	1.16020835674475\\
64	1.18004475779608\\
};

\addplot [color=BurntOrange, line width=.8]
  table[row sep=crcr]{%
1	0.148148148148148\\
2	0.148148148148148\\
3	0.444444444444445\\
4	0.444444444444445\\
5	0.444444444444445\\
6	0.444444444444445\\
7	0.444444444444445\\
8	0.444444444444445\\
9	0.444444444444445\\
10	0.444444444444445\\
11	0.740740740740741\\
12	0.740740740740741\\
13	0.740740740740741\\
14	0.740740740740741\\
15	0.740740740740741\\
16	0.740740740740741\\
17	0.740740740740741\\
18	0.740740740740741\\
19	0.740740740740741\\
20	0.740740740740741\\
21	0.740740740740741\\
22	0.740740740740741\\
23	1.03703703703704\\
24	1.03703703703704\\
25	1.03703703703704\\
26	1.03703703703704\\
27	1.03703703703704\\
28	1.03703703703704\\
29	1.03703703703704\\
30	1.03703703703704\\
31	1.03703703703704\\
32	1.03703703703704\\
33	1.03703703703704\\
34	1.03703703703704\\
35	1.03703703703704\\
36	1.03703703703704\\
37	1.03703703703704\\
38	1.03703703703704\\
39	1.33333333333333\\
40	1.33333333333333\\
41	1.33333333333333\\
42	1.33333333333333\\
43	1.33333333333333\\
44	1.33333333333333\\
45	1.33333333333333\\
46	1.33333333333333\\
47	1.33333333333333\\
48	1.33333333333333\\
49	1.33333333333333\\
50	1.33333333333333\\
51	1.33333333333333\\
52	1.33333333333333\\
53	1.33333333333333\\
54	1.33333333333333\\
55	1.33333333333333\\
56	1.33333333333333\\
57	1.33333333333333\\
58	1.33333333333333\\
59	1.33333333333333\\
60	1.33333333333333\\
61	1.33333333333333\\
62	1.33333333333333\\
63	1.33333333333333\\
64	1.33333333333333\\
};

\addplot [color=Green, line width=.8]
  table[row sep=crcr]{%
1	0.999999999999999\\
2	0.999999999999999\\
3	1\\
4	1\\
5	1\\
6	1\\
7	1\\
8	1\\
9	1\\
10	1\\
11	1\\
12	1\\
13	1\\
14	1\\
15	1\\
16	1\\
17	1\\
18	1\\
19	1\\
20	1\\
21	1\\
22	1\\
23	1\\
24	1\\
25	1\\
26	1\\
27	1\\
28	1\\
29	1\\
30	1\\
31	1\\
32	1\\
33	1\\
34	1\\
35	1\\
36	1\\
37	1\\
38	1\\
39	1\\
40	1\\
41	1\\
42	1\\
43	1\\
44	1\\
45	1\\
46	1\\
47	1\\
48	1\\
49	1\\
50	1\\
51	1\\
52	1\\
53	1\\
54	1\\
55	1\\
56	1\\
57	1\\
58	1\\
59	1\\
60	1\\
61	1\\
62	1\\
63	1\\
64	1\\
};

\addplot [color=red,line width=.8]
  table[row sep=crcr]{%
1	0.4\\
2	0.4\\
3	0.4\\
4	0.4\\
5	0.4\\
6	0.4\\
7	0.4\\
8	0.4\\
9	0.4\\
10	0.4\\
11	0.4\\
12	0.4\\
13	0.4\\
14	0.4\\
15	0.4\\
16	0.4\\
17	1\\
18	1\\
19	1\\
20	1\\
21	1\\
22	1\\
23	1\\
24	1\\
25	1\\
26	1\\
27	1\\
28	1\\
29	1\\
30	1\\
31	1\\
32	1\\
33	1\\
34	1\\
35	1\\
36	1\\
37	1\\
38	1\\
39	1\\
40	1\\
41	1\\
42	1\\
43	1\\
44	1\\
45	1\\
46	1\\
47	1\\
48	1\\
49	1.6\\
50	1.6\\
51	1.6\\
52	1.6\\
53	1.6\\
54	1.6\\
55	1.6\\
56	1.6\\
57	1.6\\
58	1.6\\
59	1.6\\
60	1.6\\
61	1.6\\
62	1.6\\
63	1.6\\
64	1.6\\
};

\node[font=\scriptsize] at (axis cs:33, 1.7) (pm8qam) {PM-8QAM};
\node[font=\scriptsize] at (axis cs:32,1.46) (w4_64) {w4\textunderscore64};
\node[font=\scriptsize] at (axis cs:11, 1.25) (4d64prs) {4D-64PRS};
\node[font=\scriptsize] at (axis cs:44,0.8) (c64) {$\pazocal{C}^*_{64}$};
\node[font=\scriptsize] at (axis cs:32,0.69) () {\textsl{x} pol.};
\node[font=\scriptsize] at (axis cs:56,0.69) () {\textsl{y} pol.};

\node[ellipse,minimum width=0.1cm,minimum height=0.2cm,draw,inner sep=0pt] (e1) at (axis cs:51, 1.6) {};
\node[ellipse,minimum width=0.1cm,minimum height=0.2cm,draw,inner sep=0pt] (e2) at (axis cs:45, 1.34) {};
\node[ellipse,minimum width=0.1cm,minimum height=0.2cm,draw,inner sep=0pt] (e3) at (axis cs:4, 1) {};
\node[ellipse,minimum width=0.1cm,minimum height=0.2cm,draw,inner sep=0pt] (e4) at (axis cs:60, 1.15) {};

\draw(pm8qam)--(e1);
\draw(w4_64)--(e2);
\draw(4d64prs)--(e3);
\draw(c64)--(e4);
\node[] at (axis cs:6, 1.6) {\textbf{(a)}};
\end{axis}

\input{tikz/OptConst64_X.tikz}
\input{tikz/OptConst64_Y.tikz}

\end{tikzpicture}%
\label{subfig:a}
\end{subfigure}%
\begin{subfigure}[b]{0.236\textwidth}
\input{tikz/Energy_levels_M=128.tikz}
\label{subfig:b}
\end{subfigure}%
\begin{subfigure}[b]{0.236\textwidth}
\input{tikz/Energy_levels_M=256.tikz}
\label{subfig:c}
\end{subfigure}%
\begin{subfigure}[b]{0.236\textwidth}
\input{tikz/Energy_levels_M=512.tikz}
\label{subfig:d}
\end{subfigure}
\vspace{-.9cm}
\caption{Energy levels of 4D constellations with unitary mean energy and spectral efficiency $m=6$ (a), $m=7$ (b), $m=8$ (c), and $m=9$ (d).}
\label{fig:energy_levels}
\vspace{-.7cm}
\end{figure}

As it can observed from the insets in Fig.~\ref{fig:energy_levels} the 2D projections over the $\textsl{x}$ and $\textsl{y}$ polarization planes of the $\pazocal{C}^*_{M}$ constellations are quite irregular. A better insight on the 4D geometry can be, however, gained by looking at the symbol energy distribution in Fig.~\ref{fig:energy_levels}. It can be seen that all $\pazocal{C}^*_{M}$ tend to polarize around two (for $\pazocal{C}^*_{64}$ and $\pazocal{C}^*_{128}$) or three (for $\pazocal{C}^*_{256}$ and $\pazocal{C}^*_{512}$) energy levels. This feature is a compromise between the constant-modulus geometry of 4D-64PRS and the large number of levels in the PM-QAM or Welti formats. This indicates that a low number of levels (but higher than one) leads to a good trade-off between linear and nonlinear shaping gain. In our optimization, this trade-off seems to be achieved by increasing the linear shaping gain whilst maintaining a fair level of nonlinearity tolerance in the constellation. This is demonstrated by the fact that all investigated constellations (for a fixed cardinality) were found to have similar optimal SNR. Only for $m=6$, $\pazocal{C}^*_{64}$, 4D-64PRS and w4\textunderscore64, achieve a more significant optimum SNR gain between 0.1 and 0.15 dB over PM-8QAM.


\vspace{-.2cm}
\section{Conclusions}
\vspace{-.2cm}
We conducted a numerical optimization of dual-polarization 4D formats in the nonlinear optical fiber channel enabled by our recently introduced 4D analytical model for nonlinear interference. Four novel 4D modulation formats were obtained, which outperform previously most known DP-4D formats at the same cardinality. This study also highlights the good performance in the nonlinear fiber channel of certain 4D formats such as the Welti constellations. Due to the likely non-convexity of the optimization problem tackled in this paper, further investigation is needed to verify whether better 4D constellations than the ones proposed here exist.

\begin{footnotesize}\vspace{.05cm}
\textbf{Acknowledgements:} {The work of G.~Liga is funded by the EuroTechPostdoc programme under the European Union’s Horizon 2020 research and innovation programme (Marie Skłodowska-Curie grant agreement No. 754462). This work has received funding from the European Research Council (ERC) under the European Union's Horizon 2020 research and innovation programme (grant agreement No. 757791).\par}
\end{footnotesize}
\vspace{-.35cm}
\bibliographystyle{osajnl.bst}
\bibliography{Biblio.bib}









\end{document}